\documentclass{ws-procs975x65}

\begin{document}
\wstoc{Vector Field Induced Chaos in  Multi-dimensional Homogeneous Cosmologies}{R. Benini, A.A. Kirillov, G. Montani}

\title{VECTOR FIELD INDUCED CHAOS IN MULTI-DIMENSIONAL HOMOGENEOUS COSMOLOGIES}

\author{R. Benini$^{1 2\dag}$, A. A. Kirillov$^{3\ddag}$, G. Montani$^{2 4 \diamond}$}

\address{$^1$Dipartimento di Fisica - Universit\`a di Bologna and INFN \\ Sezione di Bologna,
via Irnerio 46, 40126 Bologna, Italy\\
$^2$ICRA---International Center for Relativistic Astrophysics  
c/o Dipartimento di Fisica (G9) Universit\`a di Roma ``La Sapienza'',
Piazza A.Moro 5 00185 Roma, Italy\\
$^3$Institute for Applied Mathematics and Cybernetics\\ 
10 Ulyanova str., Nizhny Novgorod, 603005, Russia\\
$^4$ENEA C.R. Frascati (U.T.S. Fusione), Via Enrico Fermi 45, 00044 Frascati, Roma, Italy\\
$^\dag$\email{riccardo.benini@icra.it}
$^\ddag$\email{kirillov@unn.ac.ru}
$^\diamond$\email{montani@icra.it}}

\begin{abstract}
We show that in multidimensional gravity vector fields
completely determine
the structure and properties of singularity.
It turns out that in the
presence of a vector field the oscillatory regime
exists for any number of spatial dimensions and for all
homogeneous models.
We derive the Poincar\'e return map
associated to the Kasner indexes and fix the rules according to which the 
Kasner vectors rotate.
In correspondence to a 4-dimensional space time, the oscillatory regime here 
constructed overlap the usual Belinski-Khalatnikov-Liftshitz one.
\end{abstract}

\bodymatter

\section{Introduction}
The wide interest attracted by the homogeneous cosmological models of the 
Bianchi classification  relies over all in the allowance for their anisotropic
dynamics; among them the types VIII and IX stand because of their chaotic evolution
toward the initial singularity \cite{BKL70} that correspond to the maximum degree of generality allowed by the homogeneity constraint; as
 a consequence it was shown \cite{BKL82,K93,M95} that the generic cosmological solution can be described properly, near the Big-Bang, in terms of the homogeneous chaotic dynamics as referred to each cosmological horizon. 
However the correspondence existing between the homogeneous dynamics and the generic inhomogeneous one holds only in 
four space-time dimensions.
In fact a generic cosmological inhomogeneous model remains characterized by chaos near the Big-Bang
  up to a ten dimensional space-time \cite{DHS85,D86,EH87} while the homogeneous models show a regular (chaos free) dynamics beyond four dimensions \cite{H02,H03}.\newline
Here we address an Hamiltonian point of view showing how the homogeneous models (of each type) perform, near the singularity, an oscillatory regime in correspondence to any number of dimensions, as soon as an electromagnetic field is included in the dynamics.

\enlargethispage*{6pt}

\section{The Standard Kasner Dynamics}

Let us consider the standard $n+1$-dimensional vector-tensor theory in the ADM representation:
\begin{eqnarray}
 I=\int d^{n}xdt\left\{ \Pi ^{\alpha \beta }\frac{\partial }{\partial t} 
g_{\alpha \beta }+\pi ^{\alpha }\frac{\partial }{\partial t}A_{\alpha }
+\varphi D _{\alpha }\pi ^{\alpha }-NH_{0}-N^{\alpha }H_{\alpha
}\right\} ,  \label{act}
\end{eqnarray}
\begin{eqnarray}
 H_{0}=\frac{1}{\sqrt{g}}\left\{ \Pi _{\beta }^{\alpha }\Pi _{\alpha }^{\beta
}-\frac{1}{n-1}\left( \Pi _{\alpha }^{\alpha }\right) ^{2}+\frac{1}{2} 
g_{\alpha \beta }\pi ^{\alpha }\pi ^{\beta }+g\left( \frac{1}{4}F_{\alpha
\beta }F^{\alpha \beta }-R\right) \right\} ,  \label{hamcn}
\end{eqnarray}
\begin{equation}
H_{\alpha }=-\nabla _{\beta }\Pi _{\alpha }^{\beta }+\pi ^{\beta }F_{\alpha
\beta },  \label{momcn}
\end{equation}
Here $H_0$ and $H_\alpha$ denote respectively the super-Hamiltonian and super-momentum,
$F_{\alpha \beta }\equiv\partial _{\beta}A_{\alpha} -\partial _{\alpha }A_{\beta }$ is the electromagnetic tensor,  $g\equiv det(g_{\alpha\beta})$ is  the determinant of the n-metric, $R$ is the n-scalar of curvature  and $D_\alpha\equiv \partial_\alpha+A_\alpha$.\newline
Since the sources are absent,
it is enough to consider only the transverse components for $A_{\alpha }$ and $\pi ^{\alpha }$; therefore, we take the gauge conditions 
$\varphi =0$ and $D _{\alpha }\pi ^{\alpha }=0$.
When going over the homogeneous case, 
we choose the gauge $N=1$ and $N^{\alpha }=0$.\\
Let's adopt the Kasner parameterization,that is based on the metric 
and conjugate momentum decomposition along spatial n-bein:
\begin{equation}
g_{\alpha \beta }=\delta _{ab}l_{\alpha }^{a}l_{\beta }^{b},\,\;\ \ \ \
\;\Pi _{\alpha \beta }=p_{ab}l_{\alpha }^{a}l_{\beta }^{b},  \label{kp}
\end{equation}
We also define a dual basis
$L_{a}^{\alpha }=g^{\alpha \beta }l_{\beta }^{a}$, such that 
$L_{a}^{\alpha }l_{\alpha }^{b}=\delta_{a}^{b}$ and $L_{a}^{\alpha }l_{\beta }^{a}=\delta_{\beta}^{\alpha}$.\newline
We want to put in evidence the oscillatory regime that the bein vectors possess and so we distinguish scale functions and  the parallel from the transverse component ($\widetilde{\lambda }_{a}=\left( \pi
^{\alpha }\ell _{\alpha }^{a}\right)$)
\begin{equation}
\label{riscalate}
l_{a}=\exp \left( q^{a}/2\right) \ell _{a}\;,\hspace{1.0cm}L_{a}=\exp \left( -q^{a}/2\right) {\cal L} _{a}.
\end{equation}
\begin{equation}
\label{123}
\vec{\ell}_{a}=\vec{\ell}_{a\parallel }+\vec{\ell}_{a\perp };\;\;\;\;\;\;\vec{\ell} 
_{a\parallel }=\frac{\widetilde{\lambda }_{a}}{\pi ^{2}}\vec{\pi},\;\;\;\;\;\;\left( 
\vec{\pi}\vec{\ell}_{a\perp }\right) =0.
\end{equation}
The standard Kasner solution is obtained as soon as the limit in which all the terms $\exp \left(
q^{a}\right)$ become of higher order is taken 
\begin{equation}
\begin{array}{c}
p_{a}=const,\;\;\;\;\;\;\;\;\;\;\;\widetilde{\lambda }_{a}=const,\;\;\;\;\;\;\;\;\;\;\;\;\vec{\ell}_{a\perp
}=const, \\ 
\frac{\partial }{\partial t}q_{a}=\frac{2N}{\sqrt{g}}\left( p_{a}-\frac{1}{ 
n-1}\sum_{b}p_{b}\right) , \\ 
\sum p_{a}^{2}-\frac{1}{n-1}\left( \sum p_{a}\right) ^{2}+\frac{1}{2}\sum
e^{q_{a}}\widetilde{\lambda }_{a}^{2}=0\;,
\end{array}
\end{equation}
\begin{equation}
g_{\alpha \beta }=\sum_{a}t^{2s_{a}}\ell _{\alpha }^{a}\ell _{\beta
}^{a}\;,\;\;\;\;\;\;\;\;\;\;\;\;\;\;\;s_{a}=1-\left( n-1\right) \frac{p_{a}}{\sum_{b}p_{b}}\;,
\end{equation}
The Kasner indexes  $s_{a}$ satisfy the identities $\sum s_{a}=\sum s_{a}^{2}=1$.

\section{Billiard representation: the return map and the rotation of Kasner vectors}

If we order the $s_a$'s, the largest increasing term (as $t\rightarrow 0$
 $t^{s_{1}}\rightarrow \infty $) among the neglected ones comes from $s_{1}$  and it is to be taken into account to construct the oscillatory regime toward the cosmological singularity.
\begin{eqnarray}
\label {28}
\displaystyle\frac{\partial }{\partial t}\widetilde{\lambda }_{1} &=&0\;,\hspace{3.5cm} \frac{\partial }{\partial t}\widetilde{\lambda }_{a} =\frac{\left( \frac{ 
\partial }{\partial t}p_{1}\right) \widetilde{\lambda }_{a}}{\left(
p_{a}-p_{1}\right) }  \nonumber \\
\displaystyle\frac{\partial }{\partial t}p_{1} &=&-\frac{N}{2\sqrt{g}}\widetilde{\lambda } 
_{1}^{2}\exp \left( q^{1}\right)\;,\hspace{1.0cm}\frac{\partial }{\partial t}p_{a} =0, \\
  \nonumber \\
\displaystyle\frac{\partial }{\partial t}q_{a} &=&\frac{2N}{\sqrt{g}}\left( p_{a}-\frac{1 
}{n-1}\sum_{b}p_{b}\right) .  \nonumber
\end{eqnarray}
The first of equations (\ref{28}) gives $\widetilde{\lambda }_{1}=const$, while the second admits
the solution 
\begin{equation}
\widetilde{\lambda }_{a}\left( p_{a}-p_{1}\right) =const.
\end{equation}
The remaining part of the dynamical system allows us to determine the return map governing the replacements of Kasner epochs and the rotation of Kasner vectors ${\vec{\ell}_a}$ through these epochs 
\begin{equation}
\displaystyle s_{1}^{\prime }=\frac{-s_{1}}{1+\frac{2}{n-2}s_{1}},\;\;\;\;\;\;\;\;\;\;s_{a}^{\prime }= 
\frac{s_{a}+\frac{2}{n-2}s_{1}}{1+\frac{2}{n-2}s_{1}},
\end{equation}
\begin{equation}
\label{234}
\displaystyle\widetilde{\lambda }_{1}^{\prime }=\widetilde{\lambda }_{1},\;\;\;\;\;\;\;\;\;\;\widetilde{ 
\lambda }_{a}^{\prime }=\widetilde{\lambda }_{a}\left( 1-2\frac{\left(
n-1\right) s_{1}}{\left( n-2\right) s_{a}+ns_{1}}\right) ,
\end{equation}
\begin{eqnarray}
\label{39}
\displaystyle\vec{\ell}_{a}^{\prime }=\vec{\ell}_{a}+\sigma _{a}\vec{\ell}_{1},\;\;\;\;\;\;\;\;\sigma
_{a} =\frac{\widetilde{\lambda }_{a}^{\prime }-\widetilde{\lambda }_{a}}{ 
\widetilde{\lambda }_{1}}=-2\frac{\left( n-1\right) s_{1}}{\left( n-2\right)
s_{a}+ns_{1}}\frac{\widetilde{\lambda }_{a}}{\widetilde{\lambda }_{1}}.
\end{eqnarray}

Thus the homogeneous Universes here discussed approaches the initial singularity being described by a metric tensor with oscillating scale factors and rotating Kasner vectors.
The presence of a vector field is crucial because, independently on the considered model, it induces a  closed domain on the configuration space.

\vfill

\end{document}